\documentclass{eas}
\usepackage{graphicx}
\newcommand{\siis}{[\mbox{S\,{\sc ii}}]$\lambda$6717}
\newcommand{\siio}{[\mbox{S\,{\sc ii}}]$\lambda$6731}
\newcommand{\nii}{[\mbox{N\,{\sc ii}}]$\lambda$6584}
\newcommand{\HI}{H{\sc i}}
\begin{document}

\TitreGlobal{Mass Profiles and Shapes of Cosmological Structures}

\title{Two Dimensional Velocity Fields of\\ Low Surface Brightness Galaxies}
\author{Kuzio de Naray, R.}\address{Department of Astronomy, University of Maryland, College Park, MD, USA 20742-2421}
\author{McGaugh, S.S.}\sameaddress{1}
\author{de Blok, W.J.G.}\address{RSAA Mount Stromlo Observatory, Cotter Road, Weston Creek ACT 2611, Australia}
\author{Bosma, A.}\address{Observatoire de Marseille, 2 Place Le Verrier, 13248 Marseille Cedex 4, France}
\runningtitle{Two Dimensional Velocity Fields of LSBs }
\setcounter{page}{23}
\index{Kuzio de Naray, R.}
\index{McGaugh, S.S.}
\index{de Blok, W.J.G.}
\index{Bosma, A.}

\begin{abstract} We present high resolution two dimensional velocity fields
from integral field spectroscopy along with derived rotation curves for nine low surface brightness galaxies.  This is a positive step forward in terms of both data quality and number of objects studied.  We fit NFW and pseudo-isothermal halo models to the observations.  We find that the pseudo-isothermal halo better represents the data in most cases than the NFW halo, as the resulting concentrations are lower than would be expected for $\Lambda$CDM.
\end{abstract}

\maketitle
%

%
\section{Introduction}
Are the dark matter halos of LSB galaxies ``cuspy'' NFW halos, or are they ``cored'' isothermal halos?  NFW halos can be fit to the observations, but the cosmological parameters implied by the fits are inconsistent with the standard LCDM picture.  Isothermal halos provide much better fits, but they have no cosmological dependence or theoretical basis.  We address this question with new data that is both high resolution and two-dimensional.  

\section{Observations}
We observed 12 LSB galaxies using the DensePak Integrated Field Unit on the WIYN telescope at Kitt Peak.  DensePak is comprised of 3'' diameter fibers arranged in a 43'' $\times$ 28'' rectangle.  Velocities of the H$\alpha$, \nii, \siis, and \siio\ emission lines were measured in each fiber.  Rotation curves were derived from the two-dimensional velocity fields using the NEMO program ROTCUR.

\section{Preliminary Analysis}
We combined the DensePak rotation curves with previous longslit and \HI\  curves.  We fit the combined data with isothermal and NFW$_{free}$ halos.  We also fit an NFW$_{constrained}$ halo which was required to match the velocities at the outer radii of each galaxy, in effect, constraining the concentrations to agree with cosmology.

\begin{figure}[ht]
   \centering
   \includegraphics[scale=0.27]{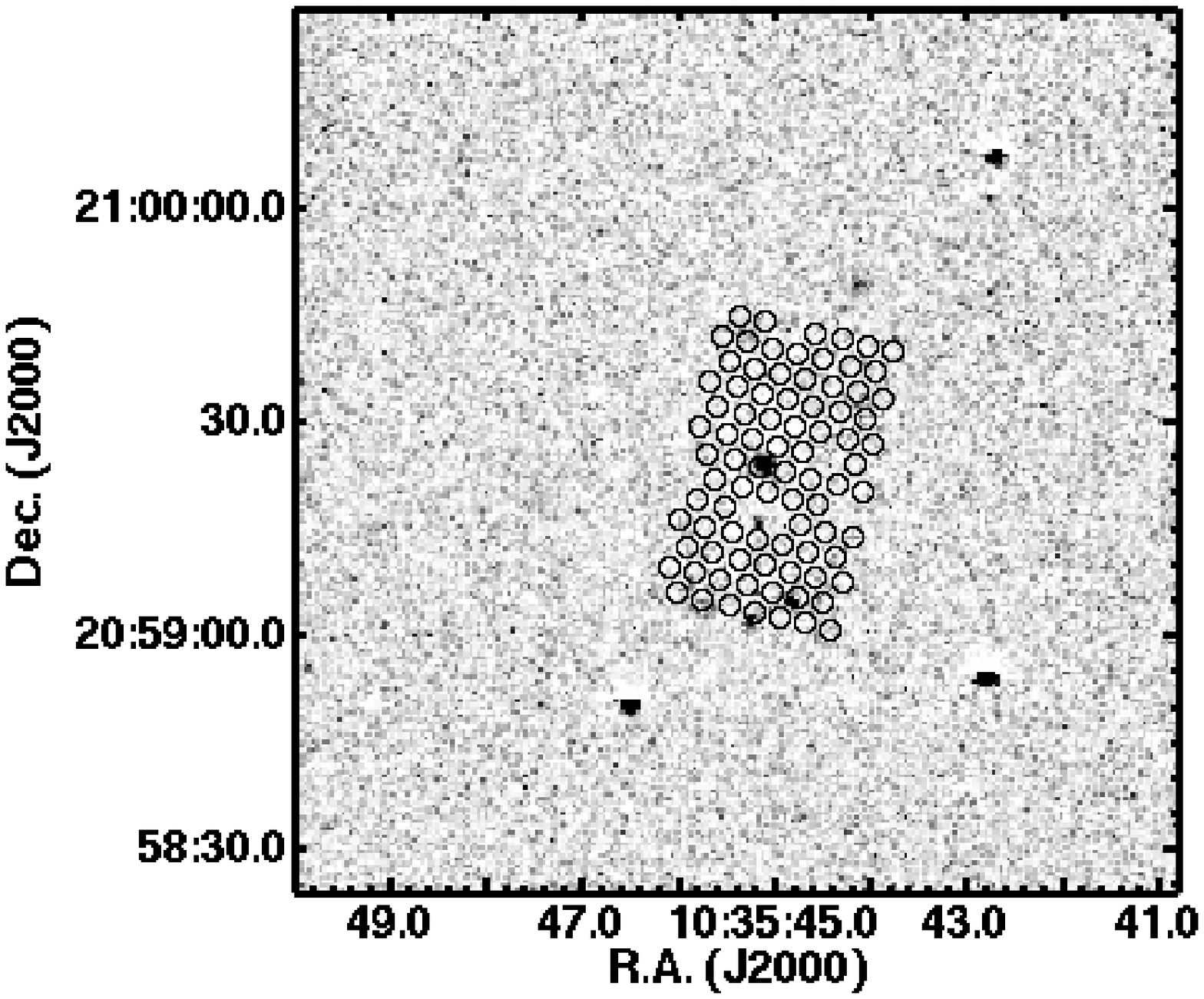}
   \hfill
   \includegraphics[scale=.23]{kuziodenaray_fig1b.ps}\\
   \includegraphics[width=4.8cm]{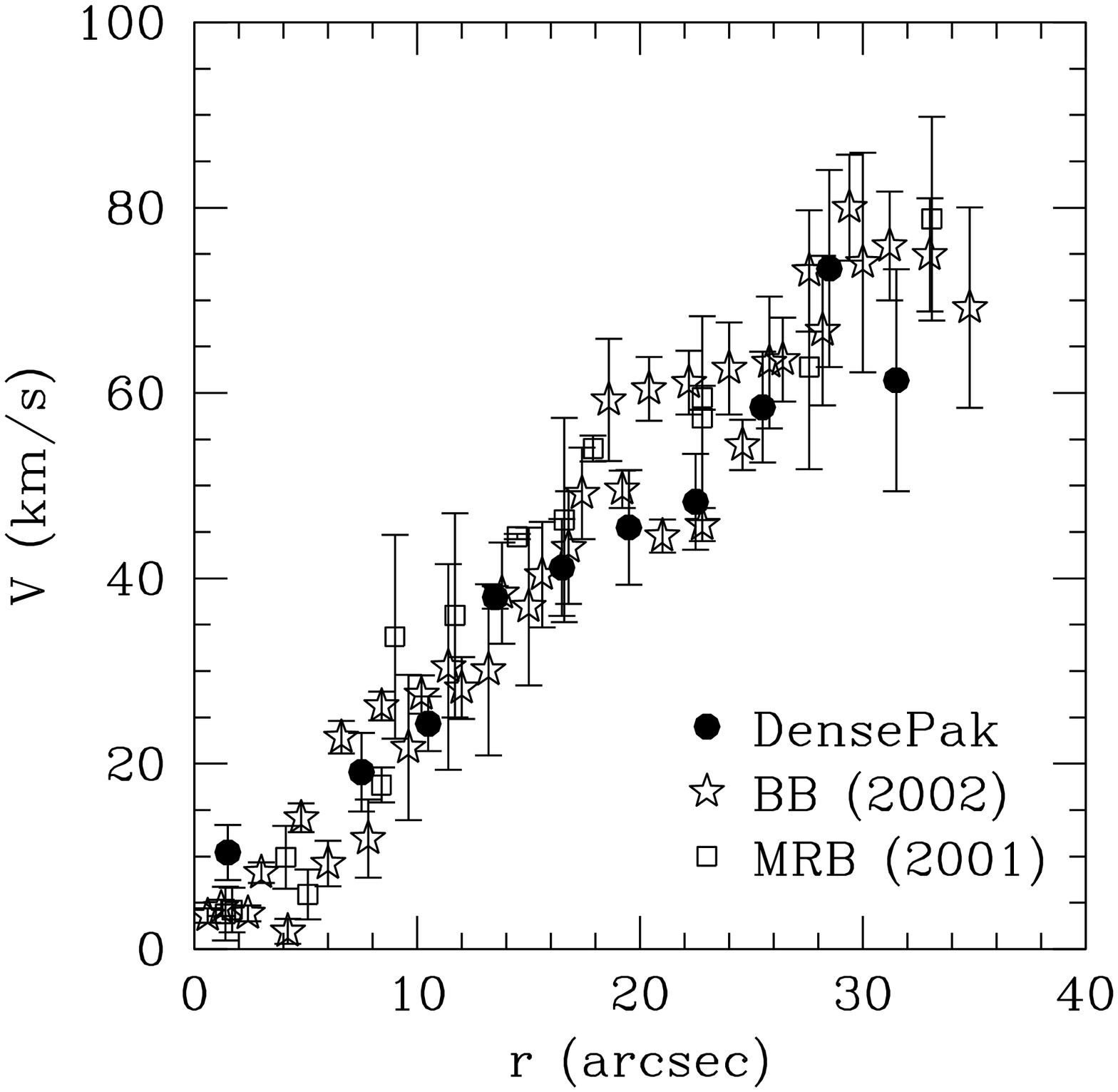}
   \hfill
   \includegraphics[width=4.8cm]{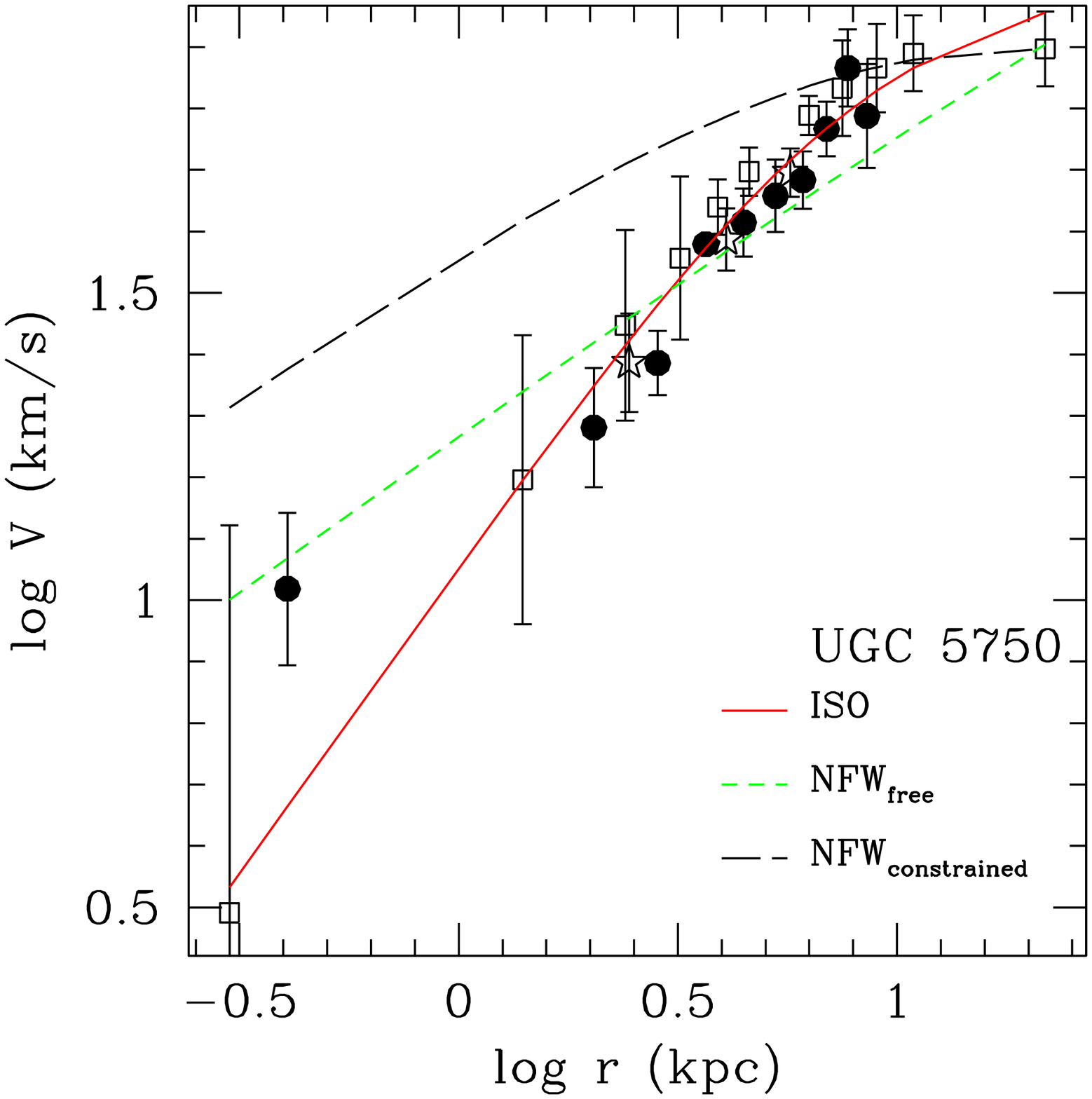}
      \caption{Clockwise from upper left: DensePak fibers on an H$\alpha$ image of UGC 5750; Observed velocity field (km s$^{-1}$); Minimum disk halo fits to combined data; Combined rotation curves.}
       \label{figure_mafig}
\end{figure}

\section{Conclusions and Future Work}
We found only one galaxy to be well fit by the NFW halo. This shows that we would have detected cusps in the other cases had they been present.  Future work will include mass modeling and an assessment of systematic effects.
\acknowledgements \textbf{Acknowledgments.} This research was supported by NSF grant AST 0206078.
\endacknowledgements

\end{document}